\documentclass[12pt,preprint]{aastex}
\usepackage[cmex10]{amsmath}
\usepackage{rotating} 		% FOR SIDEWAYS TABLE
\usepackage{url}			% FOR DISPLAYING URLS
\usepackage{subfigure}
\usepackage{graphicx}
\usepackage{times}
\usepackage{apjfonts}
\usepackage{isotope}

\shorttitle{Uranium in the H-Band}
\shortauthors{Redman et al.}

\begin{document}

\title{A High-Resolution Atlas of Uranium-Neon in the H Band}

\author{Stephen L. Redman\altaffilmark{1,2}, Gabriel G. Ycas\altaffilmark{3,4}, Ryan Terrien\altaffilmark{1,5}, Suvrath Mahadevan\altaffilmark{1,5}, Lawrence W. Ramsey\altaffilmark{1,5}, Chad F. Bender\altaffilmark{1,5}, Steven N. Osterman\altaffilmark{6}, Scott A. Diddams\altaffilmark{4}, Franklyn Quinlan\altaffilmark{4}, James E. Lawler\altaffilmark{7}, Gillian Nave\altaffilmark{2}}

\altaffiltext{1}{Department of Astronomy \& Astrophysics, The Pennsylvania State University, University Park, PA 16802, USA}
\altaffiltext{2}{Atomic Physics Division, National Institute of Standards and Technology, Gaithersburg, MD 20899, USA}
\altaffiltext{3}{Department of Physics, University of Colorado, Boulder, CO 80309, USA}
\altaffiltext{4}{Time and Frequency Division, National Institute of Standards and Technology, Boulder, CO 80305, USA}
\altaffiltext{5}{Center for Exoplanets and Habitable Worlds, The Pennsylvania State University, University Park, PA 16802, USA}
\altaffiltext{6}{Center for Astrophysics and Space Astronomy, University of Colorado, Boulder, CO 80309, USA}
\altaffiltext{7}{Department of Physics, University of Wisconsin, 1150 University Avenue, Madison, WI 53706, USA}

\begin{abstract}
We present a high-resolution (R$\approx 50~000$) atlas of a uranium-neon (U/Ne) hollow-cathode spectrum in the H-band ($1454$ nm to $1638$ nm) for the calibration of near-infrared spectrographs.  We obtained this U/Ne spectrum simultaneously with a laser-frequency comb spectrum, which we used to provide a first-order calibration to the U/Ne spectrum.  We then calibrated the U/Ne spectrum using the recently-published uranium line list of \citet{RedmanLinelist2011InPrep}, which is derived from high-resolution Fourier transform spectrometer measurements.  These two independent calibrations allowed us to easily identify emission lines in the hollow cathode lamp that do not correspond to known (classified) lines of either uranium or neon, and to compare the achievable precision of each source.  Our frequency comb precision was limited by modal noise and detector effects, while the U/Ne precision was limited primarily by the signal-to-noise ratio (S/N) of the observed emission lines and our ability to model blended lines.  The standard deviation in the dispersion solution residuals from the S/N-limited U/Ne hollow cathode lamp were $50$\% larger than the standard deviation of the dispersion solution residuals from the modal-noise-limited laser frequency comb. We advocate the use of U/Ne lamps for precision calibration of near-infrared spectrographs, and this H-band atlas makes these lamps significantly easier to use for wavelength calibration.
\end{abstract}


\section{Introduction}
The growing interest in high-precision astrophysical spectroscopy (such as the detection of low-mass exoplanets) requires the measurement of radial velocities at the meter per second (m/s) level.  This corresponds to a Doppler shift in the astrophysical spectrum of $3$ parts in $10^9$.  Such precision is being achieved now in the optical with iodine cells or hollow-cathode lamps (HCLs) with stable heavy elements such as thorium ($^{232}$Th) and a fill gas (usually argon) \citep{1992PASP..104..270M, 2009A&A...507..487M}.  In the near-infrared (NIR), sub-$10$ m/s precision calibration sources such as telluric CO$_2$ lines \citep{2010A&A...511A..55F} and gas cells \citep{2009ApJ...692.1590M, 2010ApJ...713..410B} are limited to narrow spectral regions, and thorium HCLs have few bright, calibrated lines compared to the visible. We propose that uranium neon (U/Ne) HCLs are a more suitable calibration emission source in the NIR than Th/Ar \citep{RedmanLinelist2011InPrep, ramsey2008pathfinder}.

The ideal calibration source would be a picket fence of bright emission lines or deep absorption lines covering a wide wavelength coverage and exhibiting a uniform density and uniform intensities throughout the observed spectral region.  Such a calibration source should not be too dense, because blended lines are difficult --- if not impossible --- to use for precise dispersion solutions.  The calibration source should also be repeatable over long temporal baselines, and it should be relatively inexpensive and easy to operate.

Laser-frequency combs (hereafter, LFCs) provide a picket fence of lines and are an ideal calibration source.  Even with their high cost and technical complexity, they have quickly been adopted by the astrophysical community, which is anxious for the development of such a high-precision source.  \citet{2008Sci...321.1335S} used the Vacuum Tower Telescope (VTT) solar telescope to demonstrate a comb with sunlight. Efforts are being made to develop and test such astro-combs by many groups in the optical for the High-Accuracy Radial velocity Planet Searcher (HARPS), the Trans-Atlantic Exoplanet Survey (TrES), and the High-Accuracy Radial velocity Planet Searcher North (HARPS-N) spectrographs \citep{2008Natur.452..610L, 2008SPIE.7014E..64S, 2010MNRAS.405L..16W} as well as future European Southern Observatory (ESO) facilities like the Echelle SPectrograph for Rocky Exoplanet and Stable Spectroscopic Observations (ESPRESSO) \citep{2010SPIE.7735E..14P} and the COsmic Dynamics and EXo-earth experiment (CODEX) \citep{2005Msngr.122...10P}. Our own efforts have focused on developing NIR laser-frequency combs \citep{quinlan:063105} to enable the detection of terrestrial-mass planets in the habitable zones of M dwarfs (the smallest and most populous stars in the galaxy) when coupled to the next generation of NIR precision radial velocity spectrographs like the Habitable Zone Planet Finder (HZPF) that are now being built \citep{2010SPIE.7735E.227M}. 

In the meantime, HCLs remain the next best solution for the high-precision calibration of fiber-fed spectrographs in the NIR.  Thorium is an ideal atomic emission source, as it has only one natural isotope, zero nuclear spin, a long half-life, and an abundance of lines.  Uranium shares all of these traits except the first: there are three naturally-occurring isotopes of uranium, but the abundance is dominated by $^{238}$U, which makes up $99.275~\%$ of the element, followed by $0.720~\%$ of $^{235}$U and $0.005~\%$ of $^{234}$U \footnote{\url{http://physics.nist.gov/PhysRefData/Handbook/Tables/uraniumtable1.htm}}.  Uranium also has approximately twice as many emission lines in the NIR (intrinsically).  Since only a fraction of the thorium lines in the NIR have been published \citep{kerber2008th}, a more complete uranium line list represents a significant gain in the number of measured calibration lines.  \citet{palmer1980atlas} published an atlas of uranium for the optical, and \citet{2003JQSRT..78....1E} commented on the significantly higher density of bright lines in the emission spectra of uranium compared to thorium.  There are now three uranium line lists: \citet{palmer1980atlas}, which covers the optical up through $910$ nm, \citet{RedmanLinelist2011InPrep} (hereafter, R11), which features uranium lines between $850$ nm and $4000$ nm, and \citet{1984ADNDT..31..299C}, which covers the NIR between $1800$ nm and $5500$ nm.  We refer the interested reader to the work of \citet{kerber2007future} for a review of the development of these lamps, and to \citet{kerber2008th} for a detailed NIR thorium line list.

Bright lines from the fill gases of hollow-cathode lamps are useful for moderate precision wavelength calibration, but are a hindrance for sensitive high-resolution instruments that require a very-high-precision calibration and minimal scattered light.  \citet{whaling2002argon} have demonstrated that pressure changes shift the measured wavenumbers of transitions from the $4f$, $5g$, and $6g$ levels of Ar I.  As such, these lines are not suitable for precision radial velocity measurements below tens of m/s \citep{2007A&A...468.1115L}.  In the NIR, bright argon lines hamper the analysis of the fainter Actinide lines that are more stable and suitable for wavelength calibration.  Neon, on the other hand, has relatively few lines beyond $900$ nm, and thus produces less scattered light and pixel saturation --- see figure $1$ of R11 for a comparison between Th/Ar, U/Ar, and U/Ne. These factors make U/Ne a potentially useful calibration source for precise NIR spectroscopy.

Spectral atlases are essential to the initial wavelength calibration of grating spectra.  They provide both approximate relative intensity and wavelength information that the human brain can quickly compare to an observed spectrum in a fraction of the time that would otherwise require extensive computer programming and processing time.  Atlases are useful both for gross wavelength calibration of an unknown spectral region (e.g., when a spectrograph is commissioned), and for the identification of specific spectral regions (e.g., when a telescope user obtains a spectrum at a particular grating setting).  Once the observed spectral region has been identified, the user can refer to the original line list for specific line information, and measure the line centroids to extract the dispersion solution.

We have created an atlas of uranium-neon between $1454$ nm and $1638$ nm using a simultaneous laser-frequency comb \citep{2007MNRAS.380..839M, 2007SPIE.6693E..44O, 2008EPJD...48...57B} with a second calibration fiber to provide a first-order estimate of the wavelength calibration of the hollow-cathode spectrum.  We then refined the dispersion solution using the uranium line list of R11. We observed this spectrum at a current ($14$ mA) and resolution ($50~000$) that are more typical of astrophysical spectrographs than the high-current ($75$ mA to $300$ mA), high-resolution Kitt Peak Fourier transform spectrometer (FTS) \citep{1976JOSA...66.1081B} spectra that were used to measure the uranium wavelengths in R11.  The format of this atlas makes it suitable for day-to-day use by astronomers both for visual inspection and for wavelength calibration, and its layout is comparable to Th/Ar atlases that are in routine use at telescopes around the world\footnote{\url{www.noao.edu/kpno/specatlas/thar/thar.html} and \url{het.as.utexas.edu/HET/hetweb/Instruments/HRS/hrs\_thar.html}}.

A number of stable fiber-fed instruments that are now being planned or are in the design stages could benefit from the atlas published here. The Sloan Digital Sky Survey III (SDSS-III) Apache Point Observatory Galactice Evolution Experiment (APOGEE) galactic kinematics survey is currently commissioning a $300$-fiber R$\approx$ $22~500$ H-band fiber-fed spectrograph covering the spectral region between $1530$ nm and $1680$ nm \citep{2010SPIE.7735E..46W}. New high-resolution NIR spectrographs like the Immersion Grating Echelle Spectrograph (iSHELL) \citep{2008SPIE.7014E.208T} (already under construction) can benefit from better calibration sources. The next generation of high-resolution NIR spectrographs built specifically for precision RV surveys, like the Habitable Zone Planet Finder (HZPF) \citep{2010SPIE.7735E.227M}, the Calar Alto high-Resolution search for M dwarfs with Exoearths with Near-infrared and optical \'Echelle Spectrographs (CARMENES) \citep{2010SPIE.7735E..37Q}, and SpectroPolarimetre InfraROUge (SPIRou) \citep{Donati2011InPrep}, are now in their planning stage and the U/Ne calibration source we present here may help them achieve their desired precision.

We describe the spectrometer and our H-band frequency comb in \S2. In \S3 we briefly describe how we obtained our spectra.  In \S4 we explain how we reduced and wavelength-calibrated the LFC and U/Ne data.  In \S5 we discuss some of the limitations of the atlas and compare it to the precision of the frequency comb.  This atlas is available online at {\bf URL from publisher}.

\section{Instrumentation \& Calibration}

\subsection{The Fiber-Fed NIR Pathfinder Spectrograph}

The Pathfinder testbed used for this work is similar to the instrument we used as a testbed to demonstrate $7$ m/s to $10$ m/s NIR RV precision on sunlight in \citet{ramsey2008pathfinder}. Details of our current setup, and improvements, are described in \citet{2010SPIE.7735E.231R}, but we mention key design elements here for completeness. Pathfinder is a fiber-fed warm-pupil cross-dispersed echelle spectrograph that yields a spectral resolution $\lambda/\Delta\lambda \approx 50~000$. The input fiber coupling is designed to match the f/4.2 output of the fiber from the Hobby-Eberly Telescope (HET) fiber instrument feed.  A pair of commercial NIR achromats optimized for the $0.75$ $\mu$m to $1.55$ $\mu$m region image the beam from the HET and calibration fibers onto a $100$-$\mu$m slit.  The diverging beam is folded by a small gold mirror on the way to the collimator, which in turn directs the light to a $71.5^\circ$ blaze angle (R3) echelle grating.  While this grating is undersized in the dispersion direction for the $100$-mm Pathfinder beam, it is twice as efficient as that used in \citet{ramsey2008pathfinder}. This R3 echelle operates in-plane at an off-Littrow angle of $\approx5^\circ$ to achieve high dispersion. This mode of operation (no $\gamma$ angle tilt of the echelle) ensures that there are no slit curvature effects.  A gold-coated $150$-l/mm grating blazed at $5.46^\circ$ provides cross dispersion, enabling our $1024\times1024$ HAWAII-1K array \citep{1996NewA....1..177H, 2004ASSL..300..501H} to simultaneously image $7$ to $8$ echelle orders in the Y band and $3$ to $4$ orders in the H band.   We use the same camera system as described in detail in \citet{ramsey2008pathfinder}, which uses a parabolic mirror with a weak coma-correcting lens near the entrance of the NIR dewar. The uncooled Pathfinder testbed is made possible by liquid-nitrogen-cooled ($78$ K) thermal blocking filters that suppress the out-of-band thermal radiation to a high degree \citep{MahadevanThermal2011InPrep}.

\begin{figure}[htbp]
\begin{center}
\includegraphics[width=1.0\textwidth]{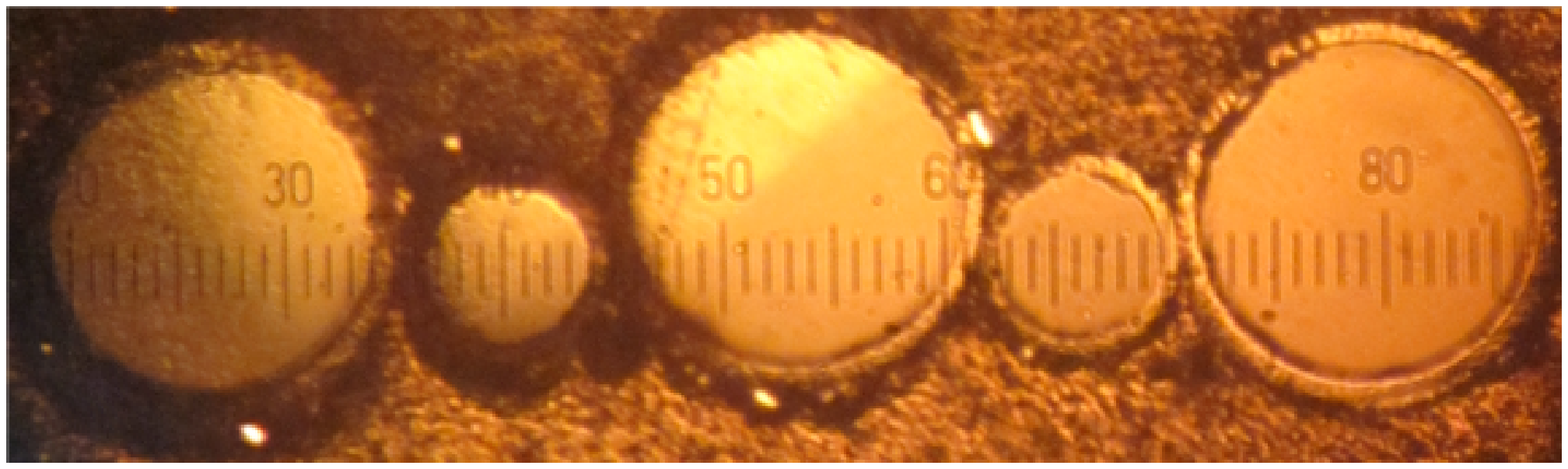}
\includegraphics[width=0.9\textwidth]{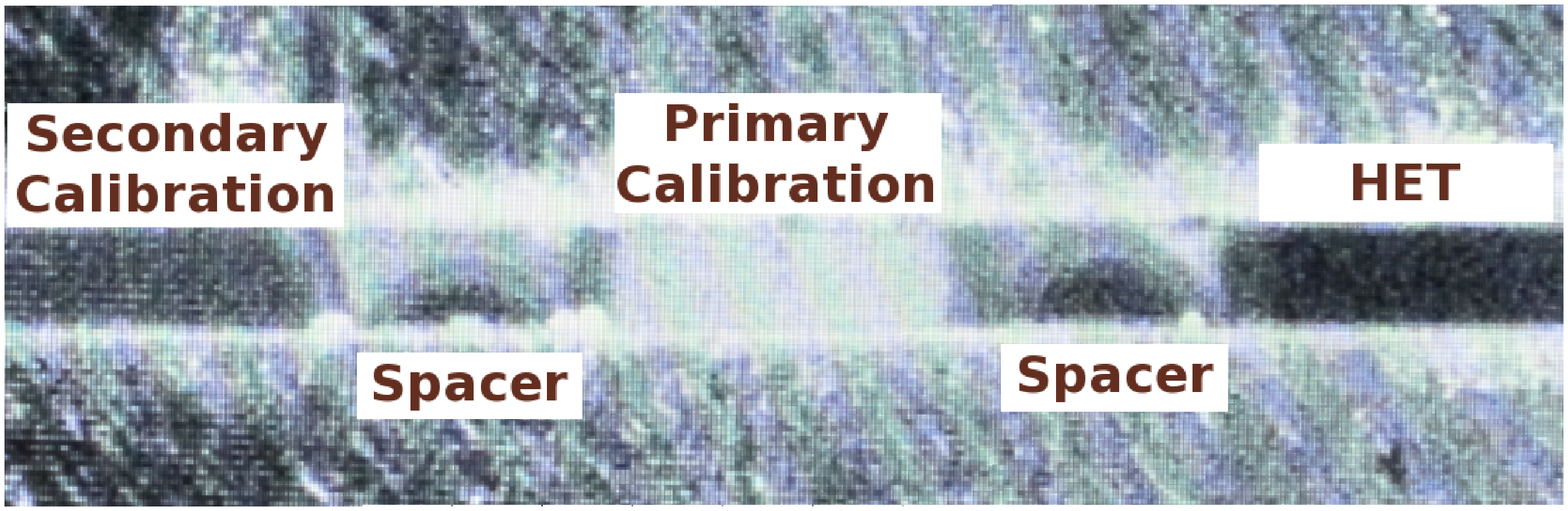}
\caption{{\bf Top:} Pathfinder Spectrograph input fiber feed.  The three $300$-$\mu$m fibers, separated by $125$-$\mu$m (outer diameter) fibers that acted as spacers for the larger fibers.  This fiber pseudo-slit was assembled and polished on-site.  From left to right, these are the secondary calibration, primary calibration, and science (stellar) fibers.  {\bf Bottom:} The same fibers, re-imaged onto a $100$-$\mu$m slit.  The slit is shown in the foreground, covering the top and bottom of the fibers.  In this image, the primary calibration fiber is transmitting light.  The slit was aligned to optimize the orientation of the primary calibration and HET fibers.  Note that the HET fiber was not used for the experiment described in this paper.}
\label{fiberslit}
\end{center}
\end{figure}

In its H-band configuration, Pathfinder has three optical fibers (one stellar fiber leading to the focal plane of the HET and two calibration fibers that lead to our calibration bench) that can simultaneously be illuminated. These fibers were polished and assembled on-site at the HET and then aligned with a $100$-$\mu$m slit, shown in Figure~\ref{fiberslit}.  The alignment of the three fibers with the slit is not perfect and this causes a constant offset in the wavelength calibration from the three fibers.  By feeding laser-frequency comb light through all three fibers simultaneously, the dispersion solution of each fiber can be independently measured to obtain the offset between the different fibers (see \S~\ref{ss_comparison}).

\subsection{The H-band Frequency Comb as a NIR Calibration Source}

\begin{figure}[htbp]
\begin{center}
\includegraphics[width=1.0\textwidth]{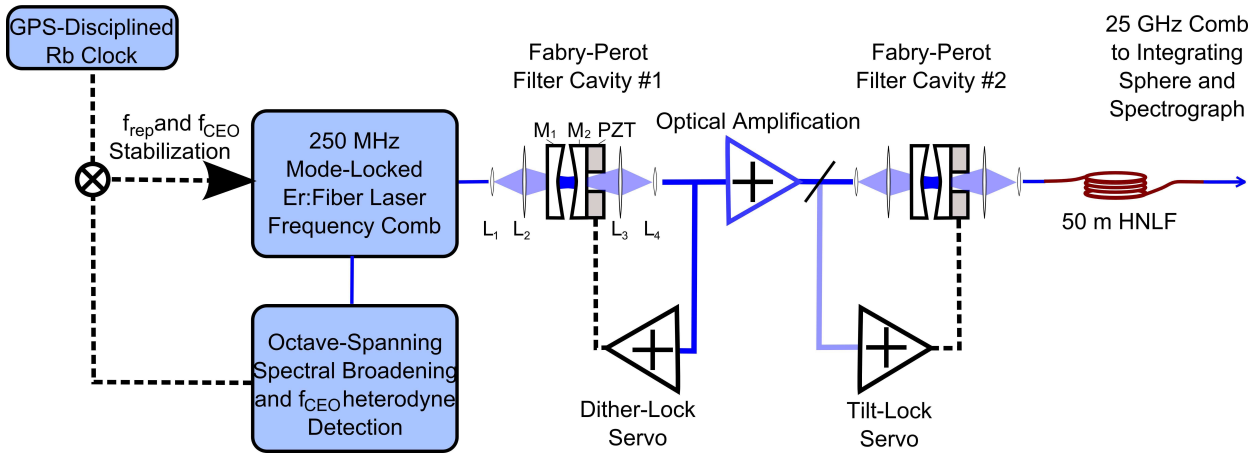}
\caption{Schematic of 25 GHz frequency comb. A 250 MHz mode-locked Er:Fiber laser is frequency controlled by locking both repetition (f$_\textrm{rep}$) and carrier-envelope offset (f$_\textrm{CEO}$)frequencies. Light from the 250 MHz frequency comb is coupled into a 25 GHz FSR Fabry-Perot cavity comprised of M$_1$ and M$_2$ with lenses L$_1$-L$_4$ matching the modes of the transmission fiber to the cavities. A piezoelectric actuator (PZT) controlled by a servo is used to actively lock the cavity length to the frequency comb. The 25 GHz comb is optically amplified, filtered again in an identical Fabry-Perot cavity, and then spectrally broadened in highly nonlinear fiber (HNLF.) Standard single-mode fiber delivers the broadened 25 GHz comb to the spectrograph, where it is launched into an integrating sphere.}
\label{asdf}
\end{center}
\end{figure}

The National Institute of Standards and Technology (NIST) and University of Colorado at Boulder (CU) H-band laser-frequency comb is built around a $250$ MHz passively mode-locked Erbium fiber laser.  The laser is frequency-stabilized by locking both the repetition rate $f_\textrm{rep}$ (direct detection) and the carrier-envelope offset frequency $f_\textrm{ceo}$ \citep[$f-2f$ optical heterodyne,][]{DavidJ.Jones04282000} to a Rb clock that is steered on the long term by the global positioning system (GPS). In this configuration, the passively mode-locked laser generates modes at frequencies of

\begin{equation}
f_\textrm{n} = n \times f_\textrm{rep} + f_\textrm{ceo},
\end{equation}

\noindent 
centered at $1~550$ nm with a bandwidth of about 70 nm. The accuracy of the mode frequencies is limited by the accuracy of the GPS-stabilized Rb clock, and is better than 1 part in $10^{11}$ \citep{quinlan:063105}.  For use with Pathfinder, the mode spacing is increased using Fabry-P\'erot filter cavities.  Two identical cavities are employed, with $25$ GHz free spectral range and finesse of about $2000$; one is placed immediately after the mode-locked laser, and the other is placed after a series of optical amplifiers, see Figure~\ref{asdf}. The lengths of the cavities are actively locked by use of piezo-electric actuators to the laser-frequency comb for maximum transmission of one set of 25 GHz-spaced comb modes. After filtering and amplification, the $25$ GHz comb is broadened in a highly nonlinear fiber, then transmitted via a single-mode fiber to an integrating sphere, where a fraction of this light enters a $300$-$\mu$m fiber that leads to the spectrograph. After filtering, the comb equation for the $25$ GHz comb is

\begin{equation}
f_n = \left(n_0 + 100 n\right) \times f_\textrm{rep} + f_\textrm{ceo},
\label{eq_fcomb}
\end{equation}

\noindent where $n_0$ is the index of one comb mode transmitted through the filter cavity, measured with a wavemeter.  For these experiments, $n_0$ was $782971$, $f_\textrm{rep}$ was $250.1132048$ MHz, and $f_\textrm{ceo}$ was $70.11320482$ MHz.  Further details of the comb and our tests with Pathfinder are presented in \citet{fcomb_ycas}.

\section{Experimental Design \& Observations}

The simultaneous frequency comb and U/Ne spectrum for this atlas were collected as part of an on-sky commissioning run of the Pathfinder spectrograph at the HET.  Light from the frequency comb was fed into the primary calibration fiber, while U/Ne light was fed simultaneously through the secondary calibration fiber (the HET stellar fiber was not needed during this experiment).  We ran a Photron\footnote{Identification of specific products is for scientific information only and does not constitute an endorsement by NIST.} U/Ne hollow-cathode lamp (Part number P863, Serial Number HGL0170) continuously at $14$ mA.  The Pathfinder HAWAII 1K detector can only cover about $20\%$ of the observable spectral range, so the grating settings had to be adjusted incrementally throughout the experiment.  At each grating setting, we obtained ten $30$-second U/Ne and frequency-comb images and five $30$-second flat-field images from an quartz tungsten halogen bulb.  All exposures were the same length so that we could take exposures continuously.  $30$-second dark frames were taken throughout the observing run and used to background-subtract the exposures.  The flat-field exposures were taken with neutral density filters in front of each fiber, which had to be removed during the U/Ne and comb exposures.  During these exchanges, a shutter in front of the fiber output in the pathfinder experiment was closed to minimize the effects of persistence in our images.  The frequency comb was monitored continuously to ensure the stability of n$_\textrm{0}$, f$_\textrm{rep}$, and f$_\textrm{ceo}$.

\begin{figure}[htbp]
\begin{center}
\includegraphics[width=1\textwidth]{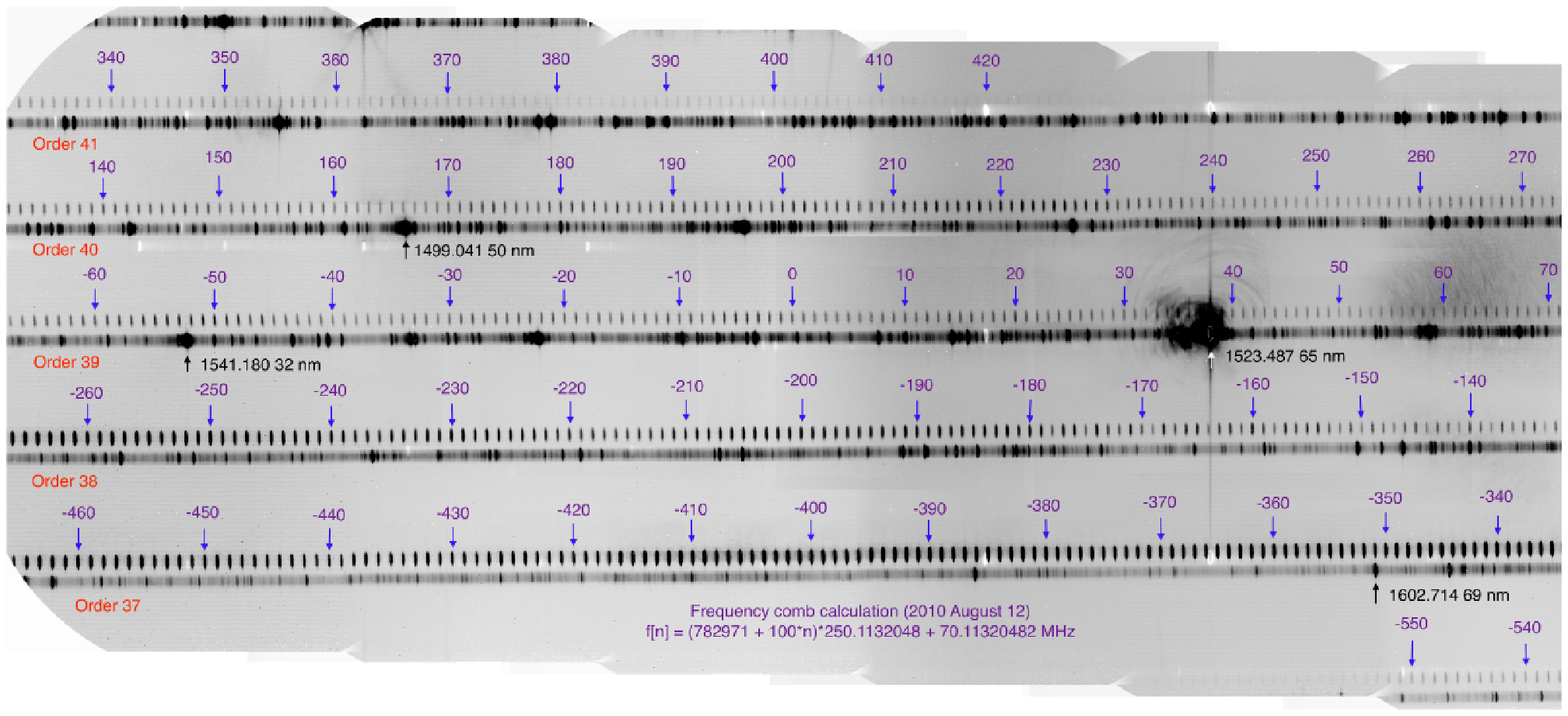}
\includegraphics[width=1\textwidth]{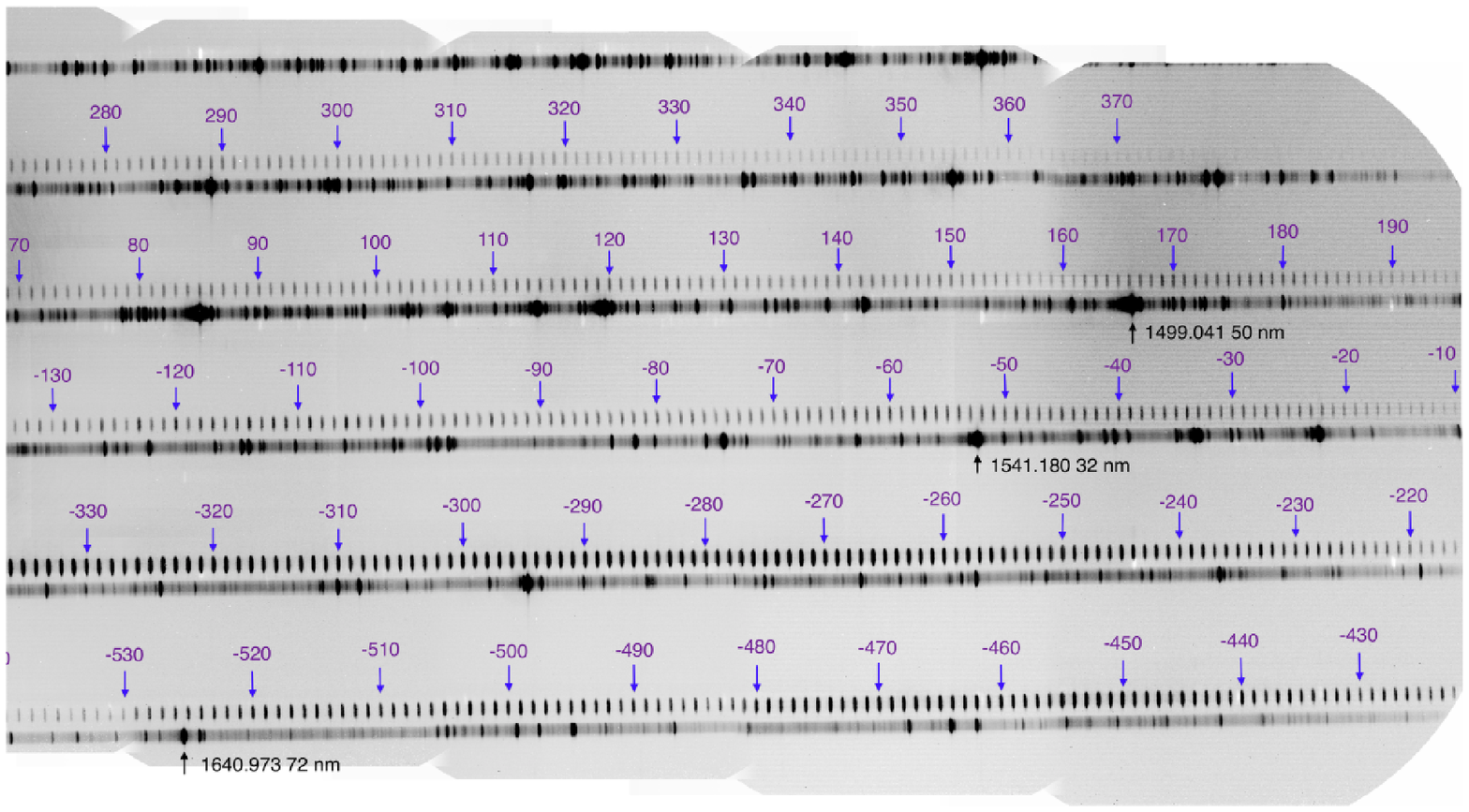}
\caption{A map of uranium-neon, taken with the Pathfinder spectrograph, and presented in two parts.  The orders are indicated in the top half of the figure.  The integers indicate the frequency comb index of the indicated line, which can be used with the frequency comb equation to calculate the frequency of any given comb line: $f_n = \left(n_0 + 100 n\right) \times f_\textrm{rep} + f_\textrm{ceo} = \left(782971 + 100 n\right) \times 250.1132048~\textrm{MHz} + 70.11320482$ MHz.  Five bright neon lines have been indicated throughout using the neon line list of \citet{saloman2004wavelengths}.}
\label{U/Nemap}
\end{center}
\end{figure}

An annotated map of U/Ne atlas, created by overlapping adjacent spectral images, can be seen in Figure~\ref{U/Nemap}.  In each order, the frequency comb spectrum is above the corresponding U/Ne spectrum.  Wavelength increases to the left and down.  The index of every tenth frequency comb line has been marked, and the frequency of each comb line can be found by using Equation~\ref{eq_fcomb}, which is also printed on the image for easy reference.  The blackbody radiation of the hollow-cathode lamp is visible as a faint continuum throughout the U/Ne spectrum; this continuum is not present in the frequency comb spectrum since the comb is not a thermal source.  Several bright neon lines have been marked throughout the map.

\section{Analysis}

\subsection{Data Reduction}
\label{ss_datared}

We constructed a median-averaged flat-field frame and a median-averaged U/Ne and laser-frequency comb image for each grating setting (in order to eliminate cosmic rays), as well as a median-averaged image of our darks to form a master dark.  This master dark frame was subtracted from each of the flat field and science frames.  We used subroutines from REDUCE \citep{2002A&A...385.1095P} to model and normalize our master flats, subtract scattered light from the U/Ne/comb images, and extract the U/Ne and comb spectra. After wavelength-calibrating the spectra (see \S~\ref{wl_cal}), the U/Ne and comb spectra at each grating setting were resampled and averaged together to produce a continuous spectrum from $1454$ nm~to $1638$ nm.  Outside of this range, the comb lines were too faint to identify.  

\subsection{Wavelength Calibration}
\label{wl_cal}

\subsubsection{Wavelength Calibration of the Frequency Comb Spectrum}
\label{ssec_wlcal_fcomb}

Wavelength identification of the frequency comb spectrum was conducted by manually identifying the comb line indices ($n$) specific to a given grating setting and fitting the comb peaks with Gaussians using MPFIT routines \citep{2009ASPC..411..251M} in Interactive Data Language (IDL).  Unlike a hollow-cathode emission spectrum, frequency comb lines are evenly spaced and identical in appearance.  Because $n_0$, $f_\textrm{rep}$, and $f_\textrm{ceo}$ are known, we can determine $n$ by feeding a single fiber with both comb light and a calibration with a well-known spectrum.  In our case, since we know the fibers are separated by less than a pixel in dispersion ($\approx 1$ GHz), while the comb lines are separated by about $25$ pixels ($\approx 25$ GHz), we used bright (but not saturated) uranium and neon lines from the adjacent fiber to determine the index of the nearest comb lines and thereby determining the index (and frequency) for all lines in that order.  Note that the determination of $n$ requires much less precision (by two orders of magnitude) than our later comparison between the FTS uranium wavelengths and the comb-based uranium wavelengths.  

The LFC dispersion solutions were found by fitting the comb line wavelengths as a function of modeled Gaussian centroids with a fourth-order polynomial, ignoring outliers that fell more than three standard deviations from the mean residual to avoid the influence of coincident cosmic rays or stray light from bright lines in adjacent orders.  An example comb spectrum can be seen in panel A of Figure~\ref{fig:disp_soln}.  The residuals of this dispersion solution can be seen in the same figure as diamonds; the scale for the residuals are on the right-hand axis.

In the absence of systematic sources of noise, the RMS of the frequency comb dispersion solution residuals should decrease as the square root of the signal-to-noise-ratio (hereafter, S/N) of the lines that were used to solve the dispersion solution.  However, we found that the frequency comb RMS residuals plateaued at about $0.000~4$ nm.  We hypothesize that this noise floor is induced by a combination of modal noise and detector inhomogeneities.

Fiber modal noise \citep{2001PASP..113..851B} is an inevitable result of the finite number of modes propagating in the fiber, and leads to additional noise in all fiber-fed spectrographs.  The modal noise in NIR fiber-fed instruments is exacerbated (compared to the visible) because the longer NIR wavelengths lead to fewer modes in the optical fiber.  In spite of mitigation efforts, like putting the LFC light through the integrating sphere and fiber agitation, we discovered that the primary and secondary calibration fibers were still impacted by modal noise.  While the effects of this were mitigated for on-sky observations by averaging together many cumulative exposures, the modal noise does affect the U/Ne and LFC calibration experiment at the level of a few tens of m/s, and sets a floor on the achievable precision.  Better agitation and scrambling mechanisms are expected to mitigate this problem.

The uncertainties of the frequency comb lines are the sum in quadrature of two components: the statistical uncertainty of the line (as measured by the width $W_{fcomb, i}$ and S/N $S/N_{fcomb, i}$ of the model Gaussian), and the uncertainty due to modal noise and detector inhomogeneities:

\begin{equation}
\delta_{fcomb, i} = \sqrt{\left(\frac{W_{fcomb, i}}{S/N_{fcomb, i}}\right)^2 + (0.000~4~\text{nm})^2}
\label{eq_fcomb_unc}
\end{equation}

\noindent Technically, we should also include the uncertainty in the frequency comb wavelength, but this is several orders of magnitude smaller than the other uncertainties and therefore negligible for our purposes.

\subsubsection{Wavelength Calibration of the Uranium-Neon Spectrum}
\label{sec:wlcal_une}

The measurement of frequency comb centroids and the calculation of the dispersion solution from their corresponding frequencies is a simple task compared to the wavelength calibration of a hollow cathode spectrum because the latter spectrum contains blended lines.  The dispersion solution of uranium neon was first estimated by offsetting the frequency-comb dispersion solution of the same grating order by $0.005$ nm, the approximate separation between the fibers.  This separation varies across the order slightly, and is different for each order and for each grating setting, so the dispersion solutions had to be solved using the U/Ne lines alone.  In an effort to derive the most accurate and precise dispersion solutions possible, we solved our U/Ne dispersion solutions in a consistent and automated manner.  The following process is applicable to the measurement of any hollow-cathode dispersion solution.

Each U/Ne spectrum was modeled as the sum of Gaussians, one for each line in the uranium and neon line lists of R11 and \citet{saloman2004wavelengths}.  For the uranium lines, we used the relative intensities of R11 to guide our initial guesses.  We also added a constant offset to the model for the continuum.  This model spectrum was iteratively modified to fit the observed spectrum using MPFIT \citep{2009ASPC..411..251M}.  We then restricted the dispersion solution to a careful selection of lines.  First, we neglected to use any neon lines for our dispersion solutions.  Some of the gas lines of argon are susceptible to pressure shifts \citep{whaling2002argon}; while it is not known whether or not any of the neon energy levels are susceptible to these shifts, we thought it prudent to ignore these lines from the beginning.  Second, we restricted ourselves to those lines with a S/N above $100$.  We lowered this threshold if fewer than $10$ lines per order had a S/N above $100$, but the threshold was usually above $40$.  Third, we used the modeled Gaussians goodness-of-fit to guide our selection.  Lines that did not appear in our observed spectra had modeled widths of zero, and lines that were poorly modeled exhibited relatively large or noisy residuals.  In an effort to avoid the influence of heavily-blended and poorly-modeled lines, we calculated the residuals as the difference between the observed spectrum and the modeled line (neglecting the other modeled lines).  If either the mean difference or standard deviation in the difference of the three closest pixels was larger than $5$\% of the line peak, the line was not used to solve the dispersion solution.  This threshold was increased in the event that there were fewer than $10$ usable lines, but was usually below $8$\%.  The uranium lines that met these criteria accounted for approximately $20$\% of the uranium lines from R11 in this spectral range, and are marked throughout the atlas with an asterisk (`*').

\begin{figure}[htbp]
\begin{center}
\includegraphics[width=0.8\textwidth]{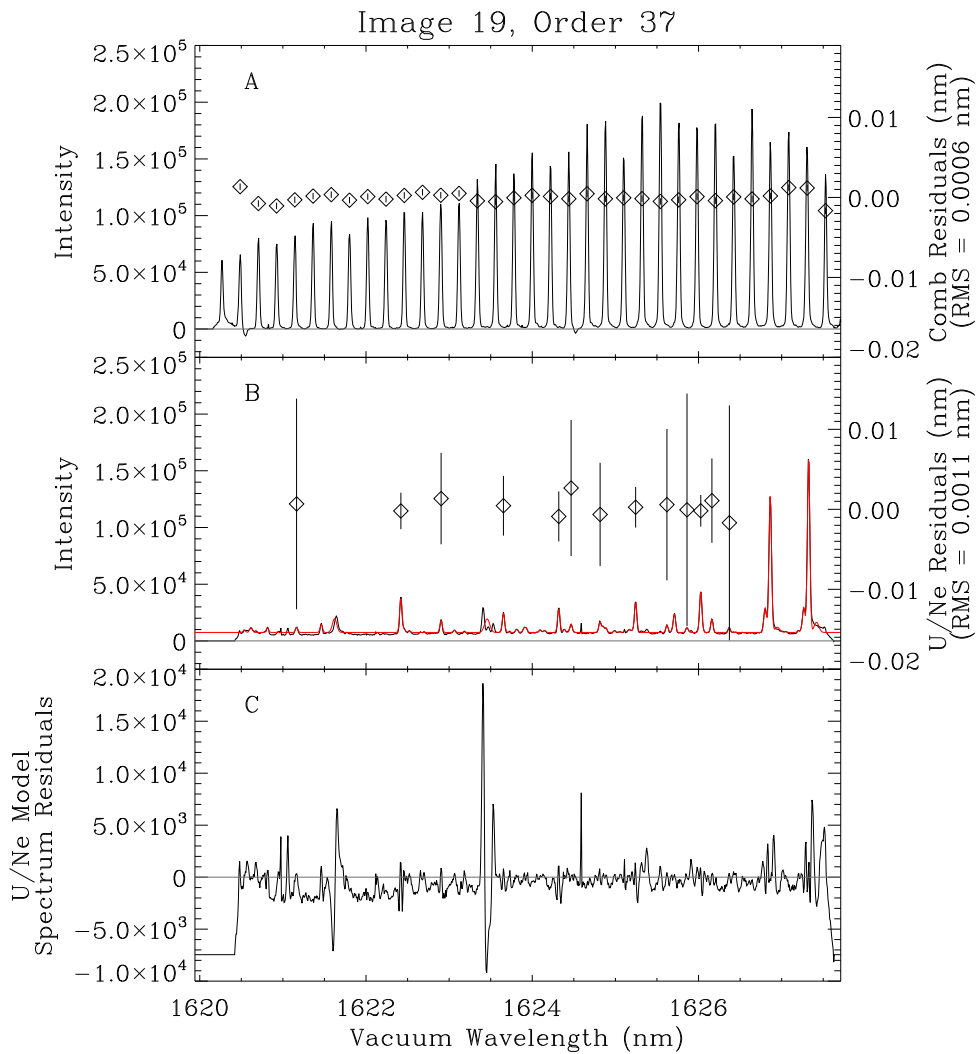}
\caption{An example comparison between the laser-frequency comb dispersion solution residuals and the uranium-neon dispersion solution residuals. {\bf A:} The centroid of each frequency comb line was found by fitting the spectral features with independent Gaussians (not shown).  The wavelength as a function of Gaussian centroid was fit with a fourth-order polynomial.  The residuals of this dispersion solution fit are over-plotted as diamonds on the frequency comb spectrum.  The scale for the residuals are on the right-hand axis. {\bf B:} The observed U/Ne spectrum was modeled as the sum of Gaussians, one for each line in the uranium line list of \citet{RedmanLinelist2011InPrep} and the neon line list of \citet{saloman2004wavelengths}.  The best-fit model, using a chi-squared minimization routine, is shown here in grey (red in electronic version).  Only certain model lines were used to solve the dispersion solutions (see \S~\ref{sec:wlcal_une} for details).  The dispersion solution was fit with a fourth-order polynomial, and the residuals are over-plotted as diamonds.  {\bf C:} The difference between the observed U/Ne spectrum and the model U/Ne spectrum.  Note that we saw several emission lines that do not correspond to known uranium or neon lines in this spectrum.  These lines (e.g., the line near $1621.7$ nm, and at least two lines near $1623.5$ nm) were not well-modeled, and show large residuals.}
\label{fig:disp_soln}
\end{center}
\end{figure}

An example uranium spectrum and model (in grey; in red in electronic version) is shown in panel B of Figure~\ref{fig:disp_soln}.  Using the above-mentioned selection of uranium lines, we solved the dispersion solutions by fitting a fourth-order polynomial to the uranium wavelengths as a function of the modeled line centroids.  The residuals to the dispersion solution are shown as diamonds.  The uncertainties of the dispersion solution residuals are the sum in quadrature of three components: the statistical uncertainty of the model line, the uncertainty due to modal noise and detector inhomogeneities, and the uncertainty of the wavelength of the line, as measured in R11:

\begin{equation}
\delta_{U/Ne, i} = \sqrt{\left(\frac{W_{U/Ne, i}}{S/N_{U/Ne, i}}\right)^2 + (0.000~4~\text{ nm})^2 + \delta_{R11_i}^2}
\label{eq_une_unc}
\end{equation}

\noindent where $W_{U/Ne,i}$ is the width of the model line and $S/N_{U/Ne,i}$ is the signal-to-noise ratio of the model line.  The residuals from the fit are shown in panel C of Figure~\ref{fig:disp_soln}.

\section{Results}
\label{sec:results}

\begin{figure}[htbp]
\begin{center}
\includegraphics[width=1\textwidth]{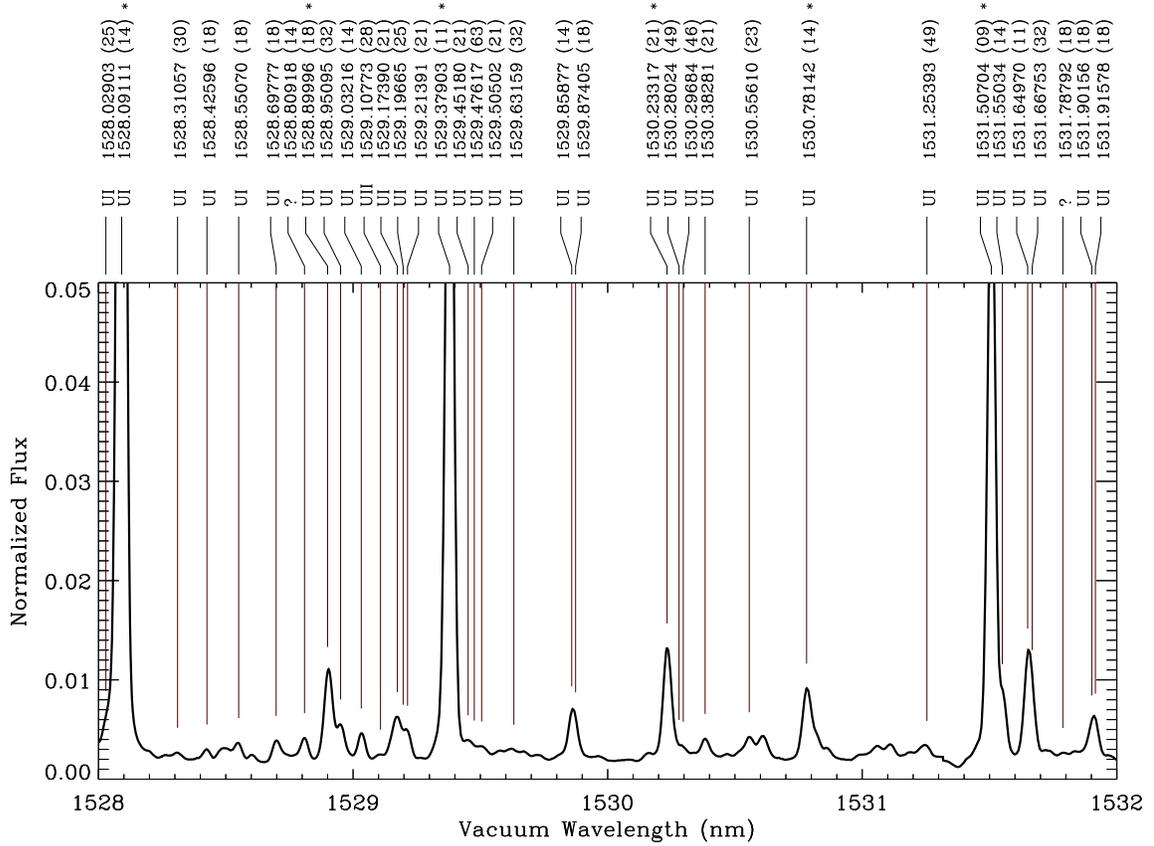}
\caption{An example image from the uranium-neon atlas.  The entire atlas is available online.  Each spectrum has been wavelength-calibrated by the brightest uranium lines.  The wavelengths and uncertainties of the uranium, unknown, and neon lines are taken from \citet{RedmanLinelist2011InPrep} and \citet{saloman2004wavelengths}.  Those lines used for the uranium-modeled dispersion solutions are marked with an asterisk (`*').  The flux has been normalized to the brightest feature in the spectrum, the (saturated) neon line at $1523.487~65$~nm.}
\label{atlas_example}
\end{center}
\end{figure}

The extracted uranium-neon H-band atlas is available online; an example of this atlas is presented in Figure~\ref{atlas_example}.  The wavelengths and uncertainties of the uranium and neon lines are taken from R11 and \citet{saloman2004wavelengths}, respectively.  Lines of an unknown origin (marked with an `?') are also from R11, and are most likely uranium lines with unknown upper energy levels.  The flux has been normalized to the brightest feature in the spectrum, the (saturated) neon line at $1523.487~65$~nm.  An electronic version of this atlas, and the wavelength-calibrated spectra used to create it are available at ({\bf URL for VIZeR, from publisher}). This atlas provides astronomers with a useful visual reference for deciding upon which spectral region to use for calibration, as well as a means of confirming the observed spectral region.

There are many features of note throughout the atlas, and we highlight a few of them here:
\begin{enumerate}
\item{There is a uranium line at $1511.052~33$ nm\space that falls within $0.006$ nm\space of a known neon line.  The uranium line is based upon its existence in U/Ar spectra.  Based upon the relative photon fluxes from R11, and the fluxes of the nearby uranium lines, both lines contribute an approximately equal amount of flux to the feature.}
\item{Some of the many blended lines are readily apparent throughout the spectrum, and confirmed both by the FTS spectra and the Pathfinder spectrum (e.g., the pair of uranium lines at $1634.1$ nm).}
\item{There are several unidentified lines throughout the spectrum, such as the line near $\approx1555.1$ nm.  These lines are not present in the uranium line list of R11, nor in the neon line list of \citet{saloman2004wavelengths}.  As such, they are not marked in the atlas, but are apparent with a visual inspection.  Others include the lines at $1578.6$ nm~and $1480.6$ nm.  These might be unknown uranium or neon lines, but they may also be due to contamination in our lamp.  No known lines of any element were found at these locations in the Atomic Spectral Database \citep{ASD4.1.0}.}
\item{Sharp artificial features sometimes appear in the spectrum when bright lines near the edge of the detector shine scattered light onto nearby grating settings.  These illumination patterns produce discontinuities in the spectral averaging algorithm; an example can be seen near the neon line at $1493.388~63$ nm.}
\end{enumerate}

\subsection{Comparison Between the Wavelength Calibrations}
\label{ss_comparison}

\begin{figure}[htbp]
\begin{center}
\includegraphics[width=0.75\textwidth]{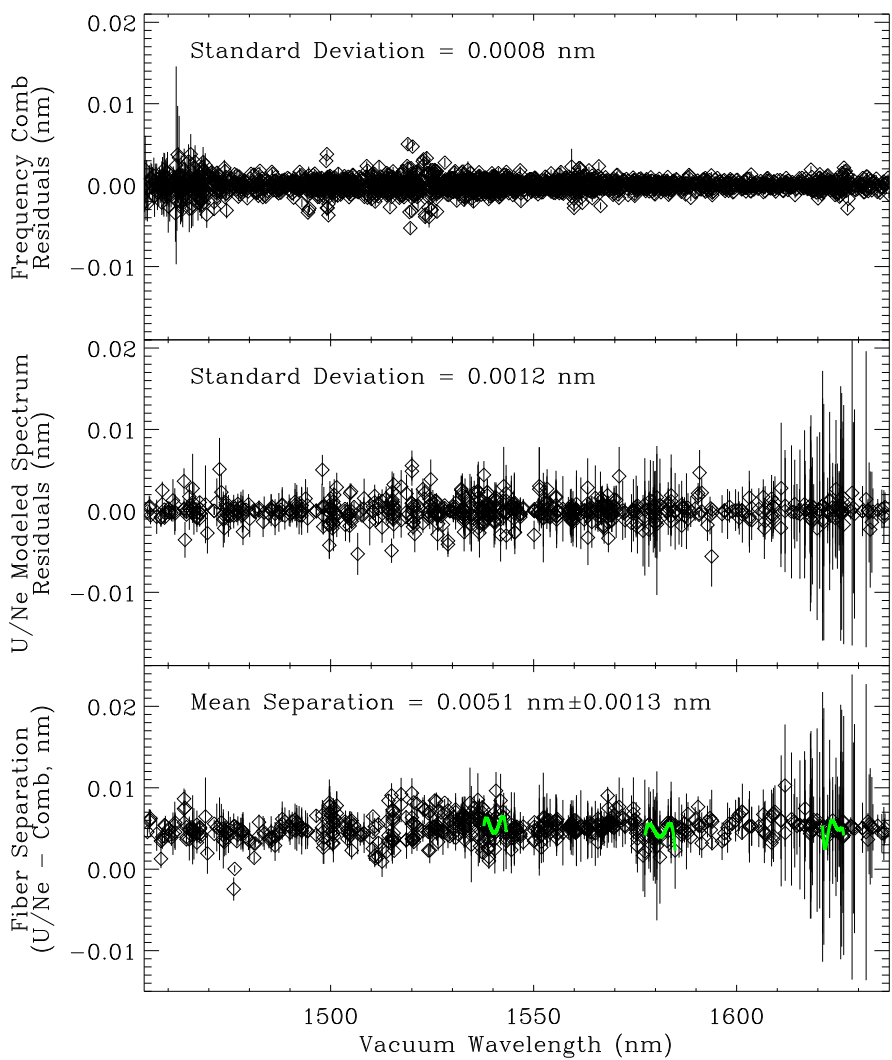}
\caption{A comparison between the precision of the frequency comb and U/Ne dispersion solutions, and an estimate of the accuracy of the U/Ne dispersion solutions. {\bf Top:} The dispersion solution residuals from fitting each comb line with a Gaussian.  The standard deviation of these points is $0.000~8$ nm.  {\bf Middle:} The dispersion solution residuals from the U/Ne spectrum, where each FTS-measured uranium and neon line has been modeled as a Gaussian and fit to the observed spectrum.  Only the brightest, best-fit uranium lines of each order were used to calculate the dispersion solutions.  This method allows us to use blended lines to a much higher degree of precision than the simpler peak-fitting algorithm.  The lines used in the bottom plot are marked throughout the atlas with an asterisk (`*').  The standard deviation of these points is $0.001~2$ nm.  The uncertainties are larger beyond about $1610$ nm~because the uranium lines in this region are weaker (in part because of a lower spectral response). {\bf Bottom:} The difference between the U/Ne and comb dispersion solutions across the H-band.  The differences presented in Figure~\ref{sep} are shown here in grey (green in electronic version).  The mean offset between the fibers is $0.005~1$ nm, and the single standard deviation of these offsets is $0.001~3$ nm.}
\label{diff}
\end{center}
\end{figure}

A comparison between the precision of the LFC and U/Ne hollow cathode dispersion solutions is shown in the top two panels of Figure~\ref{diff}.  The standard deviation of the LFC residuals was $0.000~8$ nm, while the standard deviation of the U/Ne hollow cathode residuals was $0.001~2$ nm.  The former was limited by modal noise and detector imperfections, while the latter was limited by blended lines in the spectrum, which offset the modeled centroids in a systematic fashion.

The limitation induced by modal noise was largely unexpected, as we were agitating the output end of the fiber, which is known to reduce modal noise in the visible and very near-infrared \citep{2001PASP..113..851B}.  We suspect that this modal noise was caused primarily by the way in which we fed light into the fibers: the optical alignment of the calibration system was done by eye (e.g., in the optical), and it's possible that different indices of refraction from the beam splitter caused a slight offset in the position of the H-band light relative to the optical.  By injecting fewer modes into the fiber, we diminished the number of modes at the output.  We anticipate that we can reduce modal noise in the H-band in future Pathfinder experiments by aligning our fibers in the NIR and using a double scrambler.

The uranium hollow-cathode lamp is dense with emission lines, especially in the NIR.  This density is a double-edged sword, as it provides many potential calibration lines throughout the NIR, but also a large number of blends, many of which cannot be used for precise radial velocity measurements because they obfuscate the true dispersion solution.  The precision of a given line depends upon the precise nature of the blend --- two equally bright lines will induce more of a shift in their measured line centers than will a blend where one line is significantly weaker than the other.  This is a significant advantage that a frequency comb has over hollow-cathode emission lamp sources, since blends in the former are completely avoidable, if the comb lines are optimally filtered.

Unlike measuring the precision of a wavelength calibration source, measuring its accuracy requires a comparison to wavelengths to an independent standard.  This is particularly important in the case of a hollow cathode spectrum, where the observer does not always know if there are unknown, weak lines introducing systematic errors into the dispersion solution.  This is undoubtedly the reason that the thorium-argon lamps on HARPS failed to reveal the underlying detector stitching that was so readily seen when the detector was illuminated by a LFC \citep{2010MNRAS.405L..16W}.

\begin{figure}[htbp]
\begin{center}
\includegraphics[width=0.8\textwidth]{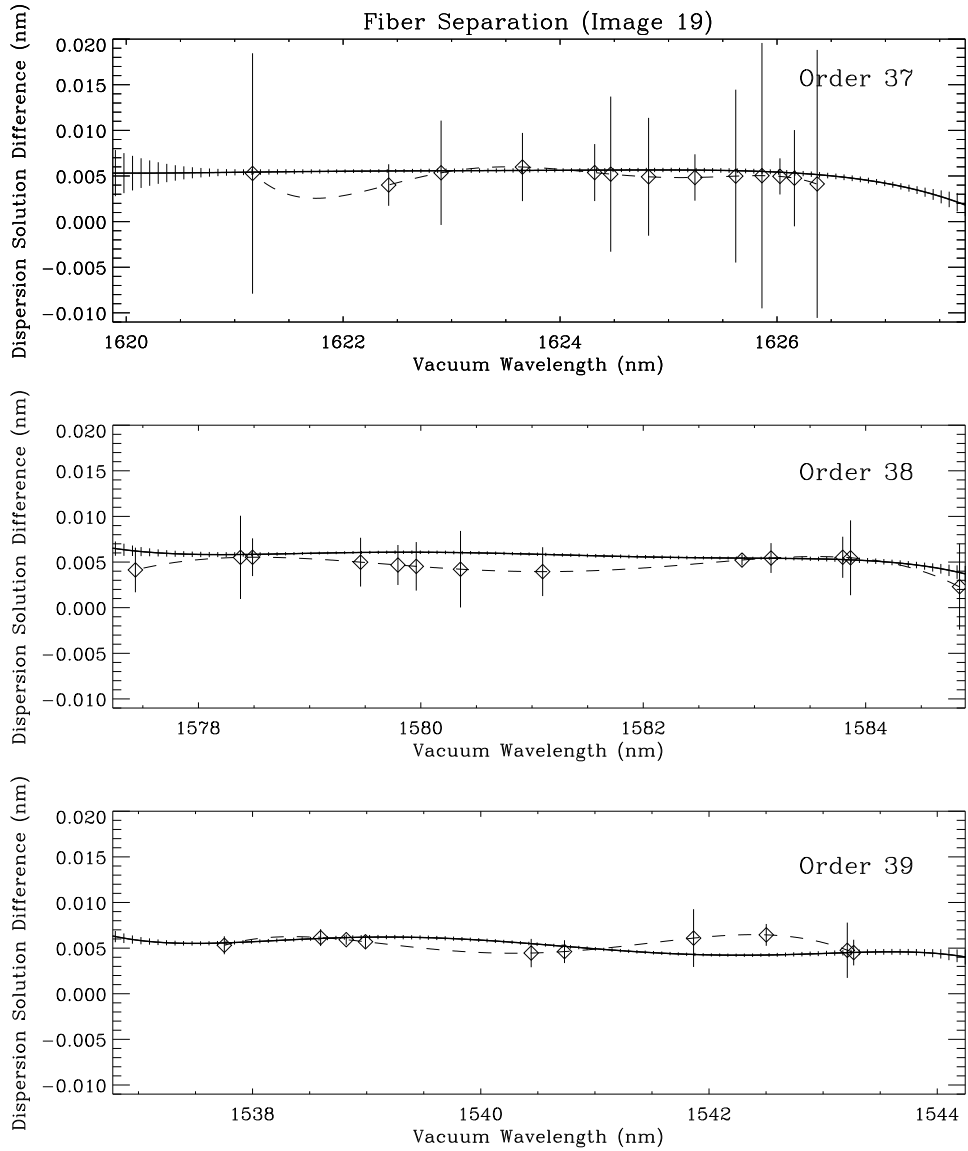}
\caption{The fiber separation between the primary calibration and secondary calibration fibers for three different echelle orders.  The solid line indicates the mean dispersion solution difference over $52$ images as measured when both fibers were simultaneously illuminated by laser frequency comb light.  The uncertainties are the single standard deviation in the dispersion solution differences over these $52$ images.  The dashed line indicates the difference between the U/Ne spectrum and the laser frequency comb over a single image.  The uncertainties on these points are the uncertainties in the uranium lines used to define the U/Ne dispersion solution.  In general, the U/Ne dispersion solutions match the dispersion solutions of the comb, within the uncertainties of the uranium lines.  Note how much the dispersion solution changes within a given order --- this is the reason that dispersion solutions from one fiber cannot be applied directly to the corresponding order of a different fiber by a constant offset, without the loss of some accuracy.}
\label{sep}
\end{center}
\end{figure}

In an effort to estimate the accuracy of our U/Ne dispersion solutions, we measured the wavelength difference between the calibration fibers by feeding them simultaneously with frequency comb light.  The dispersion solution differences between these fibers for each of three different orders is shown in Figure~\ref{sep} (solid curves).  Also plotted on this figure is the difference between the frequency comb and the U/Ne dispersion solution at the same grating setting (dashed curves).  In almost all cases, the U/Ne dispersion solution matches the frequency comb dispersion solution to within the uncertainties of the uranium lines.  Unfortunately, we could not perform this analysis at every grating angle for lack of time, but the bottom panel of Figure~\ref{diff} shows the difference between U/Ne and frequency comb dispersion solutions throughout the H-band.  The differences in the three orders of Figure~\ref{sep} are highlighted in grey (green in electronic version).  Across the H-band, the dispersion solutions exhibit approximately the expected fiber separation.

It is possible that one could achieve better precisions with a U/Ne hollow cathode source through a more careful selection of emission lines, but this experiment provides a unique opportunity to assess both the accuracy and precision of the chosen lines, which would not be possible without the LFC (or another highly-accurate and highly-precise fiducial).  Experiments at lower precision should use a more conservative selection of lines, since blending will be more frequent, while experiments at a higher precision should be able to use a larger selection of uranium lines.

\section{Conclusions}
Uranium hollow-cathode lamps have a significantly larger number of usable bright lines  than thorium hollow-cathode lamps in the NIR Y, J and H bands. This high density of lines enables precise wavelength calibration for the vast majority of NIR spectroscopic applications.  This atlas of U/Ne significantly reduces the barriers to using such lamps in modern NIR spectrographs by providing an easy-to-use visual reference.  In addition, the identification of the least-blended uranium lines enables the user to avoid regions and lines that are unsuitable.  This work also serves to illustrate the immense promise and advantage of LFCs for precision astrophysics.  While frequency combs offer the best solution for ultra-precise wavelength calibration, their cost and complexity currently limits their applications to astronomical instruments that \emph{require} an extremely high level of wavelength calibration precision over multi-year baselines. Relatively inexpensive hollow-cathode lamps are the most cost-effective approach for the majority of astronomical spectrographs.

\section{Acknowledgements}

We acknowledge support from NASA through the NAI and Origins grant NNX09AB34G, and the NSF grants AST-0906034, AST-1006676, AST-0907732, and ASF-1126413. This research was partly performed while S. Redman held a National Research Council Research Associateship Award at NIST, and while F. Quinlan was supported by a National Research Council NAS postdoctoral fellowship.  We thank M. Hirano of Sumitomo Electric Industries for use of the nonlinear fiber in the frequency comb.  This work was partially supported by funding from the Center for Exoplanets and Habitable Worlds, which in turn is supported by the Pennsylvania State University, the Eberly College of Science, and the Pennsylvania Space Grant Consortium. The Hobby-Eberly Telescope (HET) is a joint project of the University of Texas at Austin, the Pennsylvania State University, Stanford University, Ludwig-Maximilians-Universität München, and Georg-August-Universität Göttingen. The HET is named in honor of its principal benefactors, William P. Hobby and Robert E. Eberly.

We thank the HET resident astronomers (John Caldwell, Steve Odewahn, Sergey Rostopchin, and Matthew Shetrone) and telescope operators (Frank Deglman, Vicki Riley, Eusebio ``Chevo'' Terrazas, and Amy Westfall) for their expertise and support.  We also thank the HET engineers and staff who provided us with the necessary assistance during the day: Edmundo Balderrama, Randy Bryant, George Damm, James Fowler, Herman Kriel, Leo Lavender, Jerry Martin, Debbie Murphy, Robert Poenisch, Logan Schoolcraft, and Michael Ward.

\bibliographystyle{apj}

\clearpage

\end{document}